\documentclass[twocolumn]{aastex63}

\shorttitle{SED of Radio-loud quasar P352-15}
\shortauthors{Rojas-Ruiz et al.}

\usepackage{float}
\usepackage{natbib}
\usepackage{hyperref}
\usepackage[nointegrals]{wasysym}
\usepackage{ragged2e}
\usepackage{graphicx}
\usepackage{units}
\usepackage{amsmath}
\definecolor{red}{rgb}{1,0,0}
\definecolor{orange}{RGB}{204, 85, 0}
\definecolor{blue}{HTML}{4169e1}
\definecolor{ltred}{RGB}{245,167,162}
\definecolor{ltblue}{RGB}{206,211,242}

\newcommand{\gsim}{\mbox{$_>\atop^{\sim}$}}

\newcommand{\cii}{[C\,{\sc ii}]}
\newcommand{\Mdust}{\textit{M}$_{\mathrm{Dust}}$ }
\newcommand{\Lfir}{\textit{L}$_{\mathrm{FIR}}$}
\newcommand{\Ltir}{\textit{L}$_{\mathrm{TIR}}$ }
\newcommand{\angstrom}{\text{\normalfont\AA}}

\received{May 4, 2021}
\revised{July 12, 2021}
\accepted{August 1, 2021}
\submitjournal{The Astrophysical Journal}
\begin{document}

%=====================================================
%============ Title ==================================
%=====================================================

\title{The Impact of Powerful Jets on the Far-infrared Emission of an Extreme Radio Quasar at $z\sim6$}
\correspondingauthor{Sof\'ia Rojas-Ruiz}
\email{rojas@mpia.de}

%============ Name ==================================
%=====================================================
\author[0000-0003-2349-9310]{Sof\'ia Rojas-Ruiz}\altaffiliation{Fellow of the International Max Planck Research School for Astronomy and Cosmic Physics at the University of Heidelberg (IMPRS--HD)}
\affiliation{Max-Planck-Institut f\"{u}r Astronomie, K\"{o}nigstuhl 17, D-69117, Heidelberg, Germany}

\author[0000-0002-2931-7824]{Eduardo Ba\~nados}
\affiliation{Max-Planck-Institut f\"{u}r Astronomie, K\"{o}nigstuhl 17, D-69117, Heidelberg, Germany}

\author[0000-0002-9838-8191]{Marcel Neeleman}
\affiliation{Max-Planck-Institut f\"{u}r Astronomie, K\"{o}nigstuhl 17, D-69117, Heidelberg, Germany}

\author[0000-0002-7898-7664]{Thomas Connor}
\altaffiliation{NPP Fellow}
\affiliation{Jet Propulsion Laboratory, California Institute of Technology, 4800 Oak Grove Drive, Pasadena, CA 91109, USA}

\author[0000-0003-2895-6218]{Anna-Christina Eilers}\thanks{NASA Hubble Fellow}
\affiliation{MIT Kavli Institute for Astrophysics and Space Research, 77 Massachusetts Ave., Cambridge, MA 02139, USA}

\author[0000-0001-9024-8322]{Bram P.\ Venemans}
\affiliation{Leiden Observatory, Leiden University, PO Box 9513, 2300 RA, Leiden, The Netherlands}

\author[0000-0002-7220-397X]{Yana Khusanova}
\affiliation{Max-Planck-Institut f\"{u}r Astronomie, K\"{o}nigstuhl 17, D-69117, Heidelberg, Germany}

\author[0000-0001-6647-3861]{Chris Carilli}
\affiliation{National Radio Astronomy Observatory, P.O. Box O, Socorro, NM 87801, USA}

\author[0000-0002-5941-5214]{Chiara Mazzucchelli}\thanks{ESO Fellow}
\affiliation{European Southern Observatory, Alonso de Cordova 3107, Vitacura, Region Metropolitana, Chile}

\author[0000-0002-2662-8803]{Roberto Decarli}
\affiliation{INAF -- Osservatorio di Astrofisica e Scienza dello Spazio di Bologna, via Gobetti 93/3, I-40129, Bologna, Italy.}

\author[0000-0003-3168-5922]{Emmanuel Momjian}
\affiliation{National Radio Astronomy Observatory, P.O. Box O, Socorro, NM 87801, USA}

\author[0000-0001-8695-825X]{Mladen Novak}
\affiliation{Max-Planck-Institut f\"{u}r Astronomie, K\"{o}nigstuhl 17, D-69117, Heidelberg, Germany}

%=====================================================
%============ Abstract ===============================
%====================================================
\begin{abstract}
 The interactions between radio jets and the interstellar medium play a defining role for the co-evolution of central supermassive black holes and their host galaxies, but observational constraints on these feedback processes are still very limited at redshifts $z > 2$. We investigate the radio-loud quasar PSO J352.4034--15.3373 at $z \sim 6$ at the edge of the Epoch of Reionization. This quasar is among the most powerful radio emitters and the first one with direct evidence of extended radio jets ($\sim$1.6 kpc) at these high redshifts. We analyze NOEMA and ALMA millimeter data targeting the CO\,(6--5) and \cii\ far-infrared emission lines, respectively, and the underlying continuum. The broad $440\pm 80$~km~s$^{-1}$ and marginally resolved \cii\ emission line yields a systemic redshift of $z\!=\!5.832 \pm 0.001$. Additionally, we report a strong 215~MHz radio continuum detection, $88\pm 7$\,mJy, using the GMRT. This measurement significantly improves the constraints at the low-frequency end of the spectral energy distribution of this quasar. In contrast to what is typically observed in high-redshift radio-quiet quasars, we show that cold dust emission alone cannot reproduce the millimeter continuum measurements. This is evidence that the strong synchrotron emission from the quasar contributes substantially to the emission even at millimeter (far-infrared in the rest-frame) wavelengths.  This quasar is an ideal system to probe the effects of radio jets during the formation of a massive galaxy within the first Gyr of the Universe.
\end{abstract}
\keywords{cosmology: observations, reionization -  galaxies: high-redshift - quasars: individual (PSO J352.4034-15.3373)}

%====================================================
%============ Introduction ==========================
%====================================================
\section{Introduction}\label{intro}
Recent observations at high redshift ($z\gtrsim 6$, or within 1\,Gyr after the Big Bang) reveal a population of more than 300 quasars with supermassive black hole masses in the order of $\gtrsim 10^8M_{\odot}$ \citep[e.g.,][]{banados_pan-starrs1_2016,wang_exploring_2019} and host galaxies with  star formation rates (SFR) reaching up to $100-2500$ $M_{\odot}\ {\rm yr}^{-1}$ \citep[e.g.,][]{decarli_alma_2018,shao_star_2019}. The emergent picture is that the ratio of black hole mass to host galaxy mass is higher by a factor of 3--4 than expected from local relations \citep[e.g.,][]{venemans_bright_2016,neeleman_kinematics_2021}, suggesting that the black holes are growing more rapidly than their host galaxies. However, thus far all results within the first Gyr of the Universe are based solely on radio-quiet quasars,  since  there are only 12 quasars known to be strong radio emitters at this cosmic epoch, and none of their host galaxies have been studied in the millimeter (mm) thus far \citep[e.g.,][]{banados_constraining_2015,banados_discovery_2021,belladitta_first_2020, liu_constraining_2021}. Hence, the role of the radio-jets and their interaction with the interstellar medium is still unexplored at these redshifts.

Quasars at $z\sim6$ are a challenge for evolutionary models of galaxies/active galactic nuclei\,(AGN), as they have to be able to form and grow billion-solar-mass black holes in less than a Gyr \citep[e.g.,][]{inayoshi_assembly_2020}. In this context, radio jets are a possible mechanism to aid such a rapid growth by enhancing the black hole accretion rates \citep[e.g.,][]{jolley_jet-enhanced_2008,ghisellini_general_2010,ghisellini_sdss_2015,volonteri_case_2015,regan_super-eddington_2019}. Furthermore, the interplay between radio jets and the interstellar medium is thought to be an important mechanism responsible for the tight correlation between the mass of a galaxy and its central black hole \citep[e.g.,][]{kormendy_coevolution_2013}. Hydrodynamical cosmological simulations require strong AGN feedback already at $z\sim 6$ to reproduce the observed distribution of galaxy masses at $z=0$ \citep[e.g.,][]{kaviraj_horizon-agn_2017}. Observationally, however, evidence of AGN feedback has yet to be confirmed in sources at $z\gtrsim 6$ where both host galaxy and black hole are experiencing the first extreme and efficient growth \citep[see also][]{bischetti_widespread_2019,novak_no_2020}.

Studying the effects that radio jets have on the stellar properties of a galaxy is challenging because the central AGN outshines the stars \citep[e.g.,][]{zibetti_resolved_2009,taylor_galaxy_2011}. This is particularly difficult at high redshift and stellar emission from host galaxies of $z\sim 6$ quasars has yet to be detected \citep[e.g.,][]{decarli_hubble_2012,mechtley_near-infrared_2012,marshall_dark-ages_2020}. Fortunately, at these redshifts key tracers of atomic and molecular gas, such as \cii-158$\mu$m and CO, are redshifted to the sub-millimeter wavelengths and thus can be observed with facilities like the Atacama Large Millimeter/sub-millimeter Array\,(ALMA) or the IRAM NOrthern Extended Millimeter Array\,(NOEMA). \cii\ is often the brightest far-infrared (FIR) line in star-forming galaxies and can be used to infer kinematics, star-formation properties and masses of the host galaxy \citep[e.g.,][]{wang_star_2013}. The CO\,(6--5) and CO\,(7--6) molecular lines are expected to be among the strongest CO transitions in quasar host galaxies \citep[e.g.,][]{carilli_cool_2013}. Thus, the study of these tracers in radio-loud AGN can reveal whether the jets are affecting the gas for star formation in the host galaxy. 

In this paper we focus on the quasar PSO\,J352.4034--15.3373 (hereafter P352--15) at $z=5.8$, discovered in \citet{banados_powerful_2018} and confirmed to be the brightest $z\gtrsim 6$ radio object known at the time through observations with the Karl G. Jansky Very Large Array\,(VLA), with flux densities reaching up to ${\sim} 100$\,mJy at ${\sim} 200$\,MHz as measured by the Murchison Widefield Array (MWA). The only source with comparable radio luminosity at such redshift is a recently discovered  blazar at $z=6.1$ \citep{belladitta_first_2020}. Subsequent observations of P352--15 with the Very Long Baseline Array\,(VLBA) revealed the first direct evidence of extended (1.62\,kpc) radio-jets at $z\sim 6$ \citep{momjian_resolving_2018}. Recently, \citet{connor_enhanced_2021} studied this quasar with deep (265 ks) \textit{Chandra} X-ray observations, reporting an X-ray luminosity of $L_{2-10} = 1.26\times10^{45}$ erg~s$^{-1}$ and a structure aligned with the radio jets but at a distance of 50 kpc away, implying that the jets could affect larger scales than what is currently seen in radio emission (see \citealt{connor_enhanced_2021} for more details).
 
 In this work, we present the results from follow-up analysis of P352--15 with observations at millimeter wavelengths from ALMA at 290~GHz and from NOEMA at 100~GHz. We also study the radio emission at 215~MHz with observations from the Giant Metrewave Radio Telescope\,(GMRT). The data reduction is described in \S \ref{data}, followed by the analysis of the continuum and emission lines in the ALMA and NOEMA observations in \S \ref{cii-z}. We discuss the radio and  Far-infrared Radiation (FIR) properties in \S \ref{sed-text}. We present our results and comparison with previous literature in \S \ref{discussion}.  The final conclusions and discussion are presented in \S \ref{conclusion}. Throughout this work we adopt a cosmology with: H$_0$= 70 km~s$^{-1}$ Mpc$^{-1}$, $\Omega_M$= 0.3, $\Omega_\Lambda$ = 0.7, and  ${T^{z=0}_{CMB}}$= 2.725 K. Using this cosmology, the age of the Universe is 948 Myr at the redshift of P352--15 and $1"$ corresponds to 5.8 proper kpc.

%====================================================
%============ Data ==========================
%====================================================
\section{Data}\label{data}
In this section we present the reduction of the data used in this work from sub--mm to radio wavelengths using ALMA, NOEMA, and GMRT.

\begin{figure*}[pt!]
\centering
\includegraphics[width=0.52\linewidth]{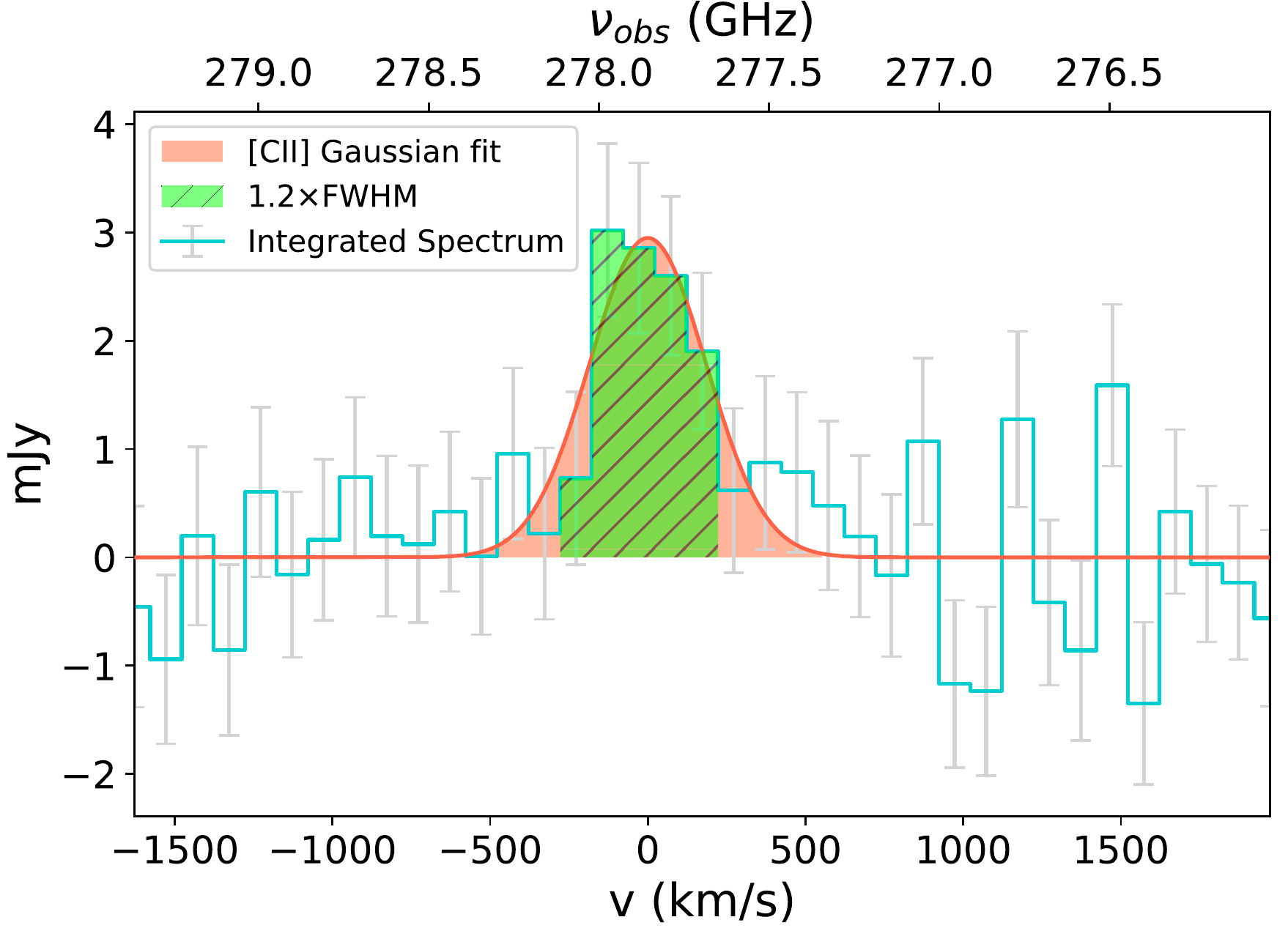}
\hfill
\includegraphics[width=0.46\linewidth]{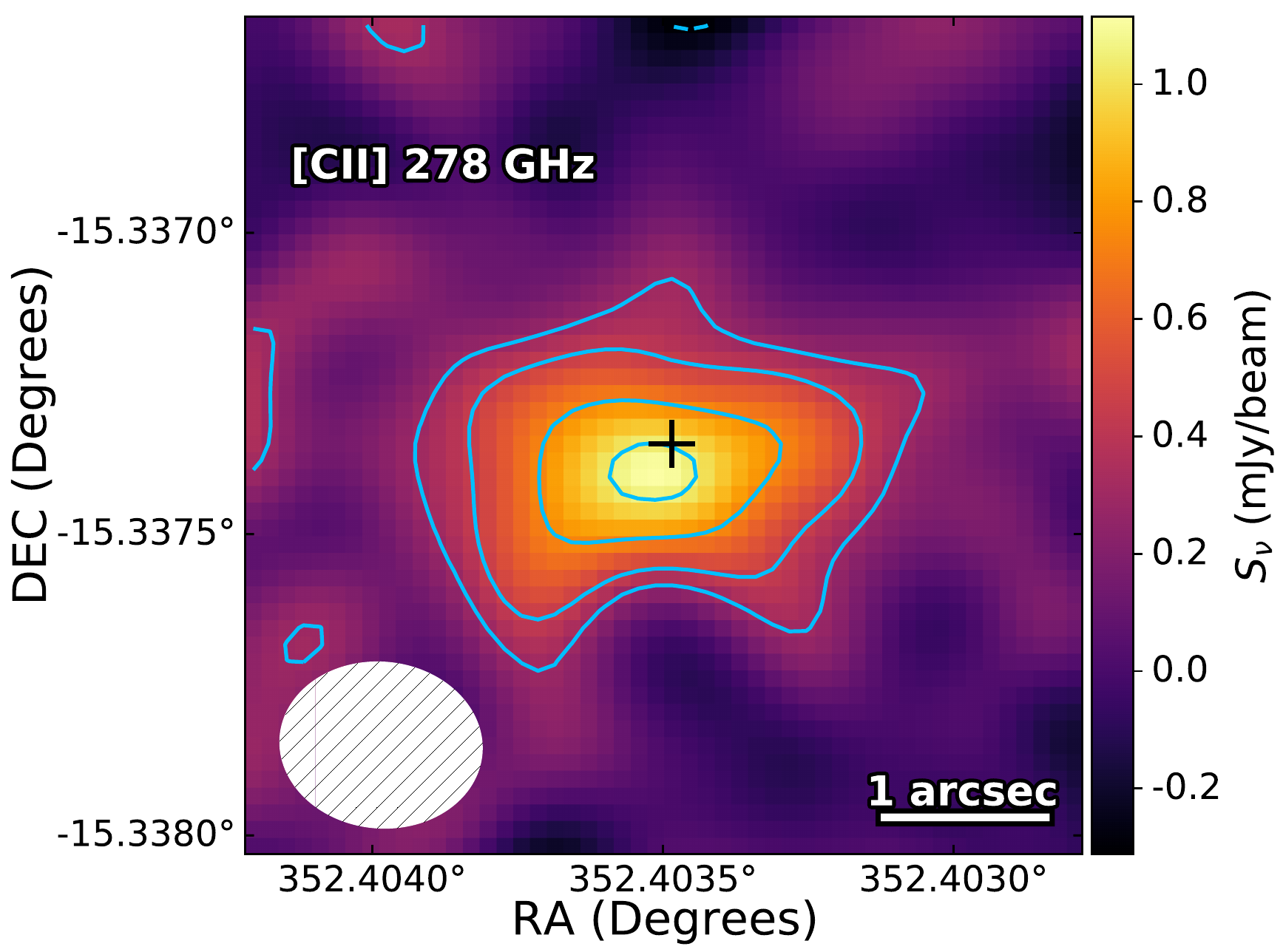}
\caption{ {\it Left:} ALMA continuum-subtracted \cii\ emission of P352--15. We use a 1\farcs7 aperture radius to extract the emission line spectra presented here in blue. The error bars represent the RMS within the aperture at each channel. The spectral range used for the Gaussian fit of \cii\ is shaded in orange and corresponds to a velocity-integrated value of $1.37\pm 0.22$ Jy km~s$^{-1}$. The \cii\ fit at peak has a flux density of $2.95 \pm 0.48$\,mJy. The hatched area in green corresponds to $\approx$ 1.2 $\times \mathrm{FWHM}$ of the Gaussian fit and is used to create the \cii\ map shown in the right panel. {\it Right:} 2D map of \cii\ built with the data from the channels shown in the left panel in green. The emission is marginally resolved and the contour levels are shown at ($-2$, $2$, $3$, $5$, $7$)$\sigma$ with $\sigma=0.15$\,mJy~beam$^{-1}$ where the Beam is $1\farcs3 \times 1\farcs$0. The \cii\ line is detected at S/N$>7.5$. The cross is centered at the optical position of P352--15 from \citet{momjian_resolving_2018}.}
\label{cii}
\end{figure*}
%====================================================
%============ ALMA ==========================
%====================================================
\subsection{ALMA}\label{alma}
The target source P352--15 was observed with ALMA on 2019 Nov 28 as part of program 2019.1.00840.S (PI: Mazzucchelli) under median weather conditions with a mean precipitable water vapor (PWV) of 1.3\,mm. The compact configuration C43-2 with a total of 44 antennas was used, resulting in maximum baselines of 314 m. Blazar J2331--1556 was used as a phase calibrator and blazar J0006--0623 was used to calibrate the flux density scale, which is accurate within 6\%. The total time on-source was 756 s. The correlator was set up to observe four spectral windows using 480 channels with a channel width of 3.9 MHz for a total bandwidth of 1.875~GHz per spectral window. Two of the spectral windows were set up to have a slight spectral overlap, and together were centered on the expected frequency of the redshifted \cii\ line. 

The data were calibrated with the ALMA pipeline, which is part of the common astronomy software application package \citep[CASA;][]{mcmullin_casa_2007}, using version Pipeline-CASA56-P1-B. A continuum image was made using the task \emph{tclean} within CASA and by applying natural weighting to optimize sensitivity. For this continuum image all channels were used except for the 0.7~GHz spectral region surrounding the \cii\ emission. The continuum image has an effective frequency of $\approx$\,290~GHz and a root-mean-square (RMS) noise sensitivity of 0.043 mJy beam$^{-1}$. We also created a continuum-free line cube by removing the continuum using the task \emph{uvcontsub}, and imaging was done with the \emph{tclean} task. Here, we again adopted natural weighting and a channel spacing of 100~km~s$^{-1}$. The RMS noise sensitivity for this cube is 0.32 mJy beam$^{-1}$ per 100 km~s$^{-1}$ channel. The final beam sizes for both the continuum image and line cube are elongated ellipses with FWHM major and minor axis $1\farcs3 \times 1\farcs0$ at a positional angle (PA) of 85.6$^{\circ}$.

%====================================================
%============ NOEMA ==========================
%====================================================
\subsection{NOEMA}\label{noema}
P352--15 NOEMA observations were executed to target the CO\,(6--5) emission line with observed frequency estimated at 101.1 GHz (program: W18EG; PIs: Eilers, Ba\~nados.) All observations were taken on thirteen different visits between 2019 April 10 and 2019 May 31. The observations were taken in the receiver Band 1 (at 3 mm) in the compact configurations of 10D with ten antennae for seven observing days, and the rest were observed in 9D configuration with nine antennae. This set up of observations were taken in compact configuration to maximize the sensitivity. Quasars 2243--123 and 2345--167 were used as phase and amplitude calibrators. The Radio Frequency bandpass calibrators included mostly 3C454.3, but also 3C84, 1749+096, and 0923+392. The star MWC 349 was used to set the absolute flux density scale, which is accurate to within 10\%. The typical system temperature was (100 -- 200) K. The precipitable water vapor conditions were mostly within ${\rm PWV}\sim 2-5\ {\rm mm}$ and with only four visits extending up to ${\rm PWV}\sim 10\ {\rm mm}$. Due to the elevation of our source, there was often shadowing, mainly in Antenna 4 and Antenna 1. The total 9 antennae equivalent time on source is 12.2 h. 

The data were reduced at IRAM with the software CLIC and MAPPING from the GILDAS suite\footnote{\url{https://www.iram.fr/IRAMFR/GILDAS/}}, and the final cube was analyzed in Python. Two cubes with resolutions of 300 km~s$^{-1}$ and 100 km~s$^{-1}$ were made to identify and analyze the CO\,(6--5) emission line and the underlying continuum. For both cubes, imaging was performed using natural weighting in order to maximize sensitivity, and cleaning was done with the simplest Hogbom algorithm. The cube with 300 km~s$^{-1}$ channel spacing has an RMS noise of 0.10 mJy~beam$^{-1}$ per channel and a synthesized beam size of $7\farcs3 \times 3\farcs8$ and PA of 5.44$^{\circ}$. The second cube has an RMS of $\sigma=0.18$\,mJy~beam$^{-1}$ per 100 km~s$^{-1}$ channel. This cube has a beam size of $7\farcs3 \times 3\farcs7$ and PA of 5.41$^{\circ}$.

%====================================================
%============ GMRT ==========================
%====================================================
\subsection{GMRT}\label{gmrt}
P352--15 was detected in the GaLactic and Extra-galactic All-sky MWA (GLEAM) survey and the TIFR GMRT Sky Survey (TGSS) at low frequencies (150 MHz to 200 MHz). However, the cataloged flux densities in this range show a wide scatter, with the GLEAM data consistent with 80 to 120 mJy at 200\,MHz \citep{hurley-walker_galactic_2017}. The TGSS image shows significant imaging artifacts (stripes), and our reanalysis of the image finds a peak of 110 mJy beam$^{-1}$, but a total flux density (significantly affected by the stripes) of 169 mJy (\citealt{intema_gmrt_2017}; \citealt{de_gasperin_radio_2018}).

To clarify the source flux density at low frequency, we observed the source with the GMRT at 215 MHz on 2018 June 23, for four hours. The observations were centered at 215.5 MHz, with a bandwidth of 25 MHz using 16384 channels, for interference excision. The flux density scale was set assuming 52.3\,Jy for 3C48 at 215\,MHz \citep{perley_accurate_2017}. Delay and bandpass calibrations were also performed using 3C48. The array gain and amplitudes were tracked in time using the calibrator
J2321--163, which had a boot-strapped flux density (from 3C48) of 14.0 Jy.
All calibrations were done using standard procedures in CASA \citep{mcmullin_casa_2007}. Interference was substantial, in particular on short baselines, and the final calibration and analysis employed only baselines longer than 1.5 km.

The final imaging used CASA task CLEAN with Briggs weighting with a robust factor$ =$ 0 \citep{briggs_imaging_1999}. The resulting synthesized beam was $= 18\arcsec \times 11\arcsec$, with PA = $-48^o$. The RMS noise on the final image is 4 mJy beam$^{-1}$.
The source was easily identified as seen in Figure \ref{cont}. Gaussian fitting indicated the source was unresolved, with a flux density of $88\pm 7$ mJy, whose uncertainty is dominated by the estimated 8\% error from the the absolute flux density bootstrap calibration process. \\

%===================================================
%============ Analysis ==========================
%====================================================
\section{Emission Line Search and mm Continuum Measurements}\label{cii-z}
In this section we present a detailed analysis of P352--15 in ALMA and NOEMA data with the purpose of measuring the \cii\ and CO\,(6--5) lines and their underlying continuum properties. 

\begin{table}
\begin{center}
\caption{Measurements from Line Emission Search \label{gaussian}}
\begin{tabular}{lccccr}
\hline
\hline
$S_{\mathrm{[C\, \sc{II}]}}$ & & & & & $2.95 \pm 0.48$ mJy\\
$S_{\mathrm{[C\, \sc{II}]}} \Delta v$ & & & & & $1.37 \pm 0.22$ Jy km~s$^{-1}$ \\
FWHM$_{[C\, \sc{II}]}$ & & & & & $440 \pm 80$ km~s$^{-1}$ \\
EW$_{\mathrm{[C\, \sc{II}]}}$ & & & & & $2.12 \pm 0.42\ \mu$m\\
$z_{\mathrm{[C\, \sc{II}]}}$  & & & & & $5.832 \pm 0.001$\\
$S_{\mathrm{CO(6-5)}}$ & & & & & $ < 0.35$ mJy\\
\hline
\end{tabular}
\end{center}
\end{table}

\subsection{\cii\ Line and 290 GHz continuum}
We use the ALMA continuum-free line cube generated in Section \ref{alma} to look for the \cii\ line. We find that the line is consistent with the expected redshift $z\sim5.84$ and looks spatially resolved to some extent. Therefore, we perform an aperture photometry test at varying apertures from 1\farcs0 to 2\farcs0. 
We perform a Gaussian fit to the \cii\ spectrum and search for the aperture at which the velocity-integrated \cii\ emission begins to plateau. This turning point occurs at a radial aperture of 1\farcs7. The resulting continuum-subtracted \cii\ spectrum is shown in Figure~\ref{cii}, which is the \cii\ spectrum used in the remaining analysis.

From the Gaussian fit, we find that \cii\ emission line has a FWHM of $440\pm 80$ km~s$^{-1}$ and the line peaks at an observed frequency of $278.20\pm 0.03$~GHz (see Figure~\ref{cii}, left). 
Using this value and the rest frame frequency of \cii\  \citep[1900.5369 GHz;][]{schoier_atomic_2005}, we calculate the systemic redshift of P352--15 to be $z\!=\!5.832 \pm 0.001$. The flux density peak is $2.95\pm 0.48$ mJy and the velocity-integrated line is $1.37\pm 0.22$ Jy km~s$^{-1}$. To generate a \cii\ map, we re-imaged the data cube using channels within 1.2$\times \mathrm{FWHM}$ of the \cii\ emission line which, given our observations, corresponds to the green shaded channels in Figure \ref{cii}. Using this channel spacing maximizes the signal-to-noise on the emission line and recovers 84$\%$ of the flux emission, assuming it is Gaussian \citep[see Appendix A in][]{novak_no_2020}. The \cii\ map is shown in the right panel of Figure~\ref{cii} and has an RMS of 0.15 mJy beam$^{-1}$. We note that  the beam size is $1\farcs3 \times 1\farcs0$ and therefore does not resolve the radio jet revealed by the VLBA, which extends over $0\farcs28$ (see \citealt{momjian_resolving_2018}). 

%######## Velocity Maps Figure
\begin{figure}%[tb]
\centering
\includegraphics[width=\linewidth]{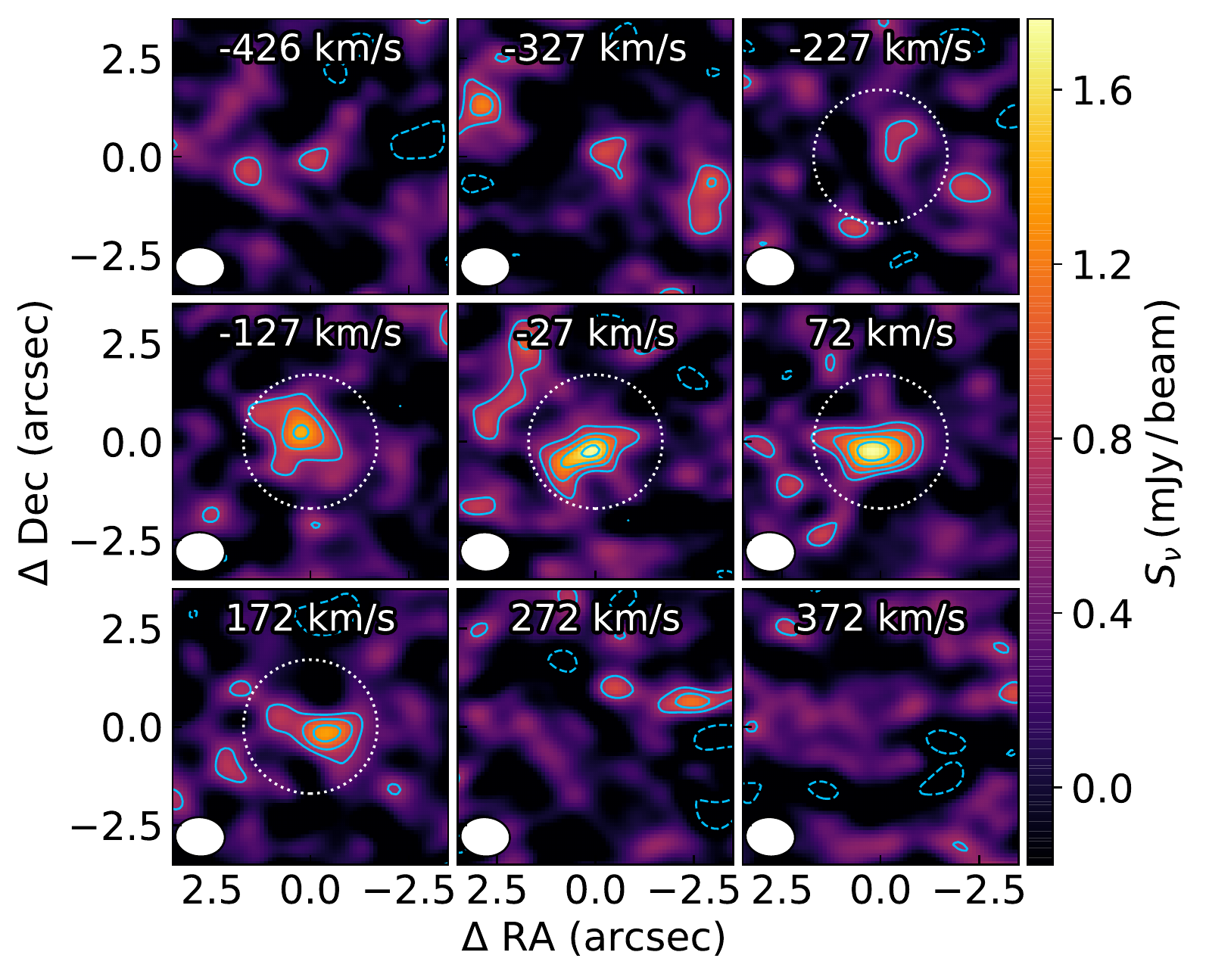}

\caption{ALMA 278\,GHz \cii\ emission in 100 km~s$^{-1}$ channel maps of P352--15. The central map corresponds to the channel closest to the peak velocity at -372.57 $\pm$ 34.91~km~s$^{-1}$ from the fitted \cii\ emission model (orange zone in Figure \ref{cii}). For each panel, the central coordinates correspond to R.A. (J2000) $= 23^\mathrm{h} 29^\mathrm{m} 36\fs8363$, DEC. (J2000) $= -15^{o}20'14\farcs460$, the channel velocities relative to the the peak velocity are given at the top, and the synthesized beams are shown at the bottom left. The channels used for flux extraction are presented with a white dotted circle of 1\farcs7 aperture radius. In all maps the contour levels are shown at ($-2$, $2$, $3$, $4$, $5$)$\sigma$ with their respective $\sigma$ values $\sim 0.03$ mJy~beam$^{-1}$.}
\label{vel_maps}
\end{figure}

The \cii\ map shows an extended morphology that we fit with a 2D Gaussian profile using CASA. The result is a deconvolved source with major and minor axes of sizes $2\farcs0 \times 0\farcs8$ and position angle $96^\circ$, confirming that the \cii\ emission of the quasar host galaxy is marginally resolved. Thus, we decide to investigate its morphology from channel-to-channel. Figure~\ref{vel_maps} shows the ALMA channel maps for P352--15 around the observed frequency of \cii, with the dotted white circles denoting the channels and spatial regions used to extract the \cii\ flux emission and create the 2D map shown in Figure \ref{cii}. The central panel shows the \cii\ emission closest to the peak velocity from the Gaussian fit ($-$372.57 $\pm$ 34.91~km~s$^{-1}$) at $-27$ km~s$^{-1}$ away. This channel and the $72$ km~s$^{-1}$ centered channel (middle right panel) have the highest signal-to-noise ratio with S/N$>5$. At the spatial resolution and S/N of our observations there are no significant changes in morphology per channel. The current data are insufficient to draw any conclusions on galaxy dynamics. Higher spatial resolution and S/N observations are needed to investigate the true velocity structure of the \cii\ emission line in this quasar host galaxy.

Finally, we utilize the reduced ALMA continuum cube (refer to Section \ref{alma}), to measure the underlying continuum at 290 GHz. The emission is unresolved with a flux density of $0.34\pm 0.04$~mJy (see Figure \ref{cont}, top).

\subsection{CO (6--5) Line and 100 GHz continuum}
\begin{figure}[ht!]
\centering
\includegraphics[width=0.91\linewidth]{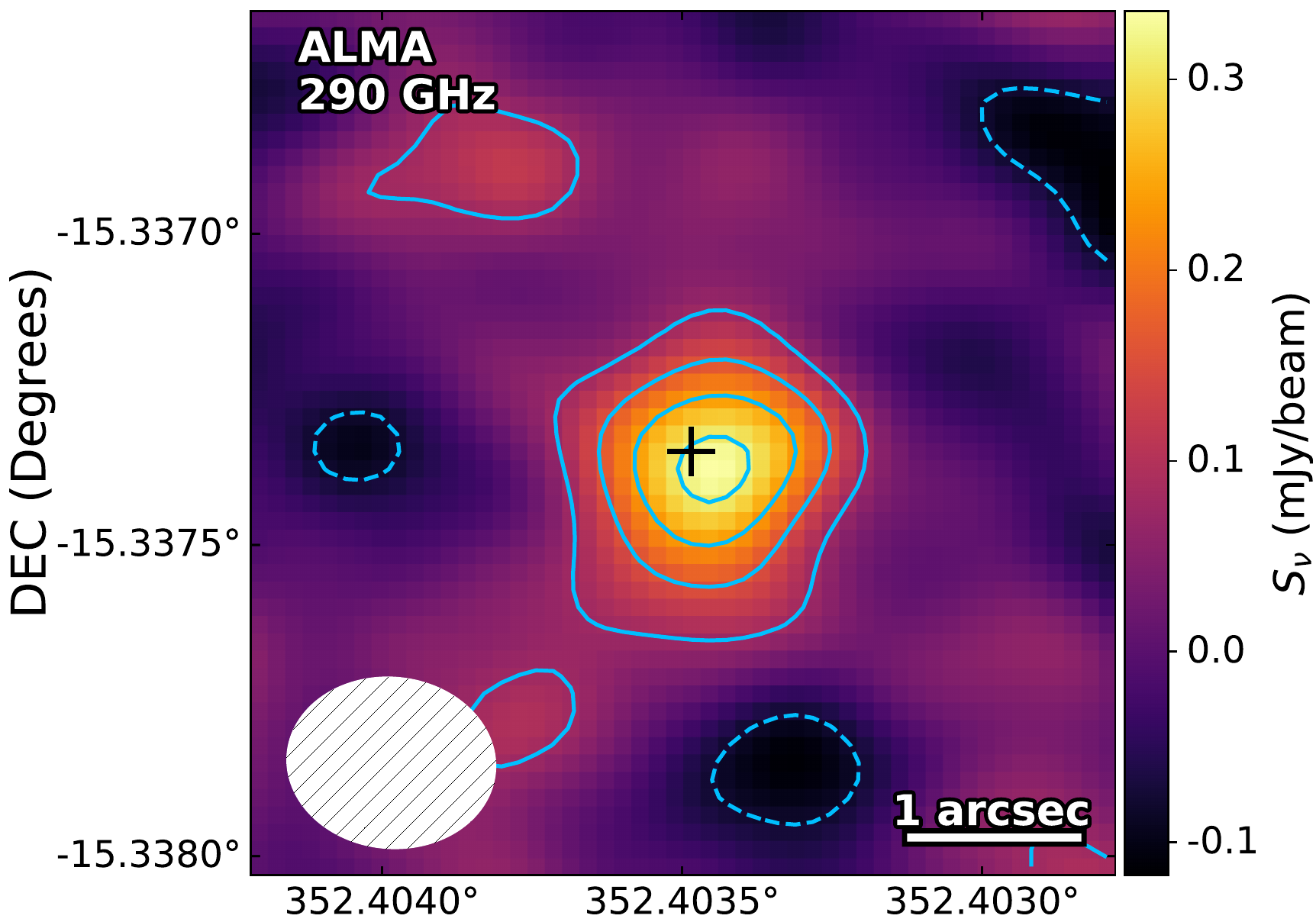}
\includegraphics[width=0.91\linewidth]{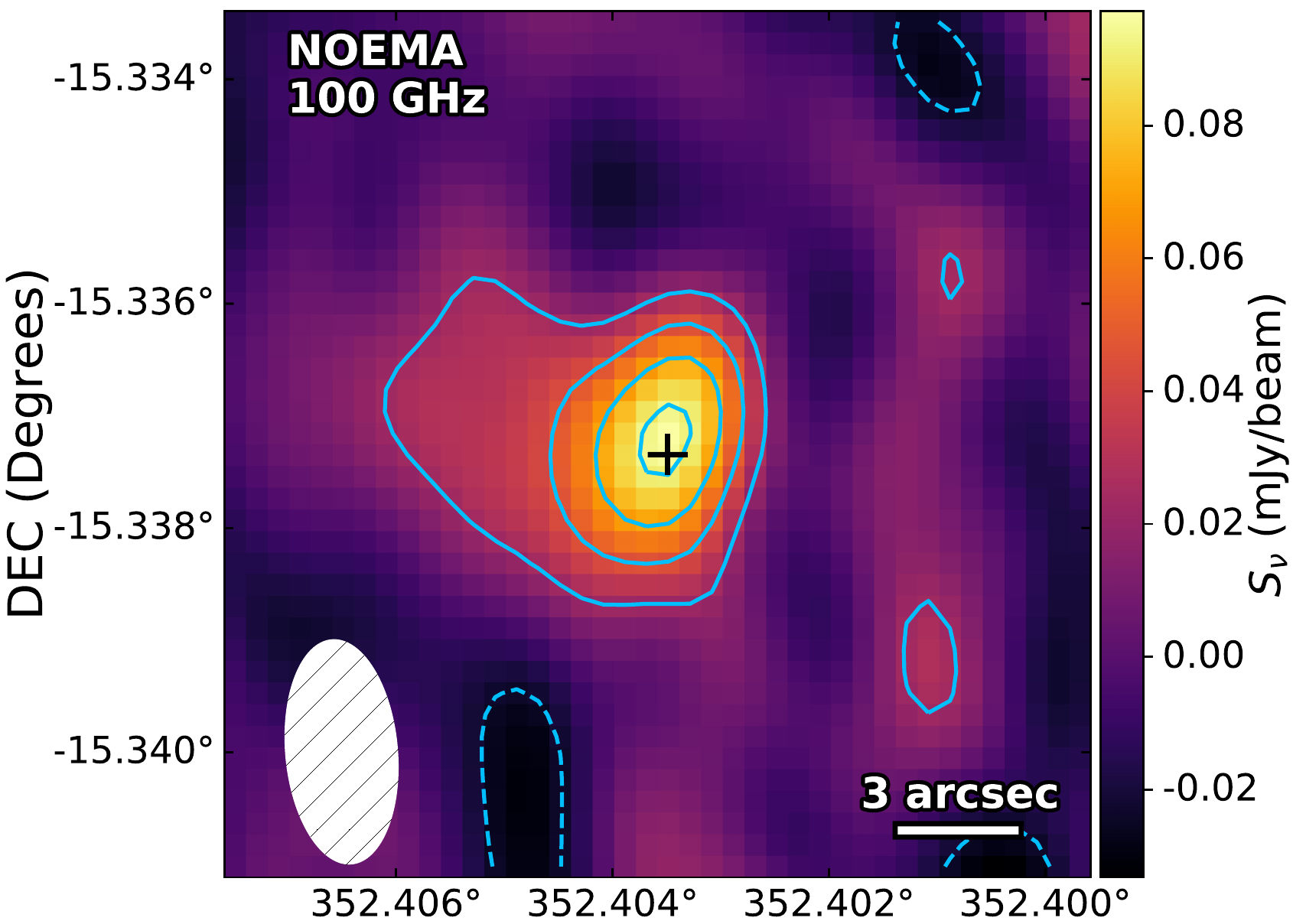}
\includegraphics[width=0.91\linewidth]{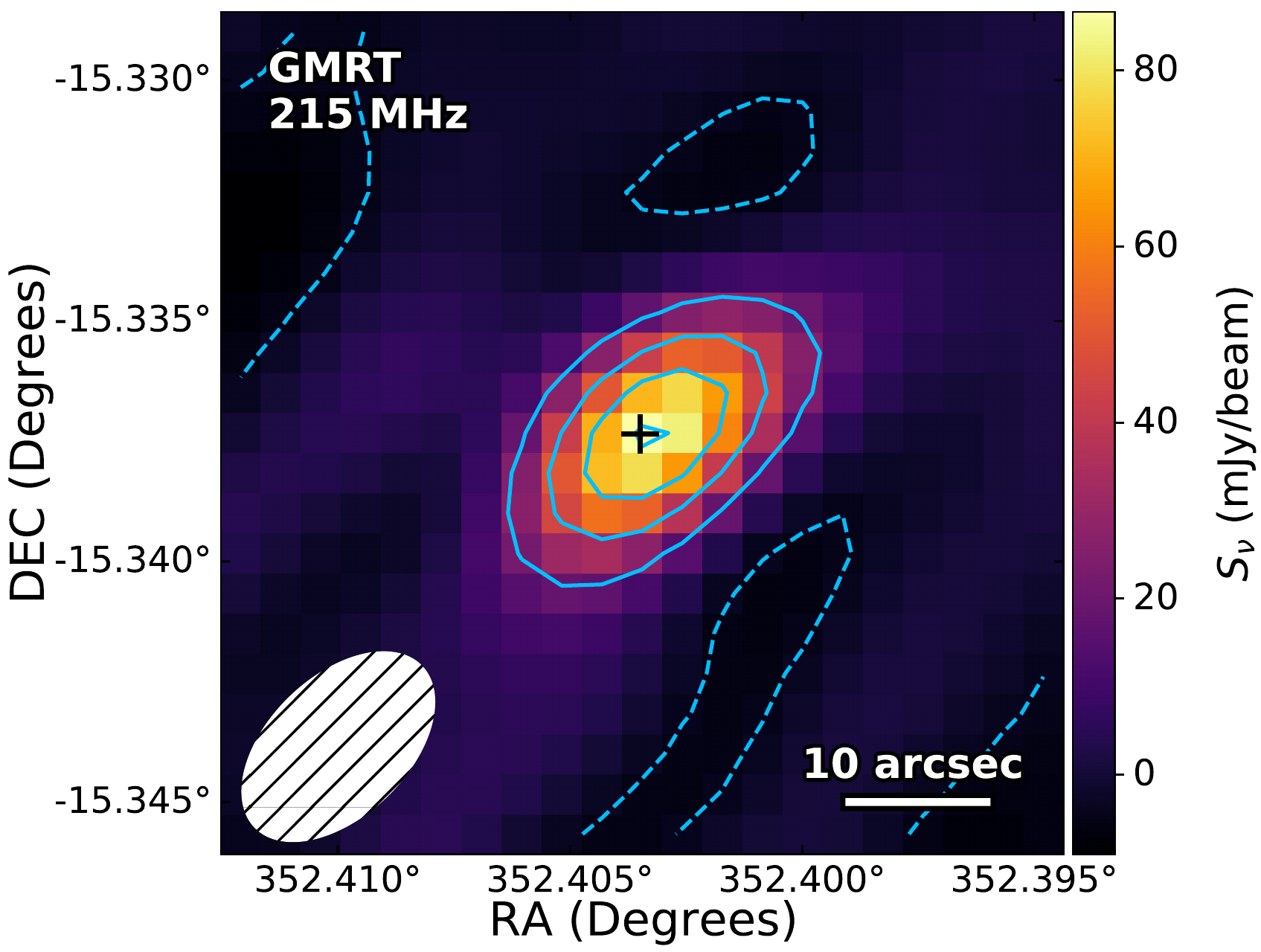}
\caption{{\it Top:} ALMA 290\,GHz continuum emission map of P352--15, with an RMS $\sigma = 0.04$\,mJy beam$^{-1}$. The beam size is $1\farcs3 \times 1\farcs0$ and the quasar is unresolved. {\it Middle:} NOEMA 100\,GHz continuum emission map with $\sigma = 0.01$\,mJy beam$^{-1}$. The beam size is $7\farcs3 \times 3\farcs7$ and the quasar is unresolved. {\it Bottom:} GMRT 215\,MHz emission map with a beam size of $18\arcsec \times 11\arcsec$ and $\sigma = 4$ mJy beam$^{-1}$. The quasar appears unresolved. The contour levels for ALMA and NOEMA maps are shown at ($-2$, $2$, $4$, $6$, $8$)$\sigma$ and both have S/N$>8$. The GMRT continuum has contours shown at ($-1$, $5$, $10$, $15$, $20$)$\sigma$ and has S/N$>20$. For all maps, the dashed and solid contours represent the negative and positive $\sigma$, respectively. The cross in all panels is centered at the optical position of P352--15 from \citet{momjian_resolving_2018}.}
\label{cont}
\end{figure}

We use the re-sampled NOEMA cubes at 300 km~s$^{-1}$ and 100 km~s$^{-1}$ described in Section \ref{noema} to search for the CO\,(6--5) molecular emission line. We do not find a significant detection of the line in either of the cubes. In order to estimate an upper limit on the CO\,(6--5) luminosity, we use the cube with the 100 km~s$^{-1}$ velocity increment and assume that the line has the same FWHM as the measured \cii\ emission line. We find a 3-$\sigma$ upper limit of 0.35 mJy, which corresponds to $L_{\mathrm{CO(6-5)}} < 0.5 \times 10^9$ $L_\odot$. This upper limit is comparable to the median values for CO\,(6--5) measured in $z>6$ radio-quiet quasars \citep[e.g.,][]{venemans_molecular_2017}. 
The 100\,GHz continuum emission is clearly detected and unresolved with a flux density of $0.10 \pm 0.01$\,mJy (Figure~\ref{cont}).

%====================================================
%============ SED ==========================
%====================================================
\section{The Spectral Energy Distribution of P352--15}\label{sed-text}
 
We build the spectral energy distribution (SED) for P352--15 by making use of the data presented in this paper and previously reported optical data from Gemini/GMOS $r^\prime$, Pan-STARRS1 \textit{i, z, y} bands and Magellan/FourStar $J$ band, and radio observations with the VLA at 3.0 and 1.4\,GHz \citep[see][]{banados_powerful_2018, connor_enhanced_2021}. In the following subsections, we will investigate the FIR properties of the quasar and will refine its radio properties. We summarize all existing measurements in Table \ref{photometry} and Figure \ref{sed}.

%%%%%%%%%%%%%%%%vertical continuum maps

\subsection{Modeling the Radio Emission}\label{model-radio}
The radio-continuum spectrum is dominated by synchrotron emission at rest-frame frequencies $\lesssim 10$ GHz, where the thermal bremsstrahlung (free-free) emission is weak \citep{duric_separation_1988}. Free-free emission becomes significant at higher frequencies which for this quasar results in a negligible contribution ($\ll 1\mu$Jy) for the study of the SED  \citep[e.g.,][]{venemans_copious_2017,yun_radiofarinfrared_2002}. As shown in Figure \ref{sed}, the synchrotron radiation of P352--15 is well described as a simple power law of the form $S_{\nu} \propto \nu^{\alpha}$, where $S_{\nu}$ is the observed flux density at the frequency $\nu$ and $\alpha$ is the radio spectral index. In \citet{banados_powerful_2018} the synchrotron power-law slope was not well constrained because of the large scatter at the lower-frequency observations. \citet{banados_powerful_2018} assumed two cases for their analysis with radio slope index $\alpha^{0.150}_{1.4} = -0.89$ and $-1.06.$ With our higher S/N GMRT data (see Section \ref{gmrt}), we can obtain a more robust measurement of the synchrotron radio slope. We fit a power law to the data from observed frequencies 3~GHz, 1.4 GHz, and 215 MHz, resulting in a radio spectral index of $\alpha^{0.215}_{3} = -0.88 \pm 0.08$. 
 
%%%%%%%% Compute Radio loudness
The new spectral index presented here allows for an improved calculation of the radio-loudness of P352--15. We adopt the definitions for radio-loudness, for which a  quasar is considered radio-loud when 
$R_{4400} = f_{\nu,\, \mathrm{5\, GHz}}\, /\, f_{\nu,\, \mathrm{4400\,\angstrom}} >10$ or $R_{2500} = f_{\nu,\, \mathrm{5\, GHz}}\, /\, f_{\nu,\, \mathrm{2500\,\angstrom}}>10$ \citep{sramek_radio_1980,kellermann_vla_1989}. From our power-law fit we calculate a flux density of $3.87 \pm 0.10$ mJy at rest-frame 5 GHz. We use the L$_{\mathrm{4400\,\angstrom}}$ reported in \citet{banados_powerful_2018} and convert it to flux density and find $f_{\nu,\, \mathrm{4400\,\angstrom}} = (3.53 \pm 0.89)\times 10^{-3}$ mJy. We therefore calculate the radio-loudness $R_{4400} = 1100 \pm 280$, which agrees with the $R_{4400}\gtrsim 1000$ reported in \citet{banados_powerful_2018}. Similarly, from the  L$_{\mathrm{2500\,\angstrom}}$ we calculate $f_{\nu,\, \mathrm{2500\,\angstrom}} = (2.63^{+0.19}_{-0.16})\times 10^{-3}$\,mJy, and measure $R_{2500} = 1470^{+110}_{-100}$. 
%\edit1{
Although the radio-loudness of  P352--15 is extreme and similar to those observed in blazars \citep[e.g.,][]{belladitta_first_2020,belladitta_extremely_2019,sbarrato_sdss_2012,romani_q09066930_2004}, its X-ray properties \citep{connor_enhanced_2021} confirm its quasar nature. The parameters of the X-ray-to-optical index $\alpha_{\mathrm{ox}} = -1.45 \pm 0.11$, and the photon index $\Gamma = 1.99^{+0.29}_{-0.28}$ fall under the quasar classification although near the threshold of high-redshift blazars from \citet{ighina_x-ray_2019} where a blazar has $\tilde{\alpha}_{\mathrm{ox}} <1.355$\footnote{$\tilde{\alpha}_{\mathrm{ox}} = 0.789 \alpha_{\mathrm{ox}} + 0.212(\Gamma - 1.0)$} and $\Gamma < 1.8$. Typically, the photon index is $\Gamma \geq 1.5$ for high-redshift quasars even up to $z \gsim 6$ \citep[e.g.,][]{nanni_500_2018}. Additionally, the radio jet orientation of P352--15 not pointing along our line of sight \citep{momjian_resolving_2018} and the steep radio slope ($\alpha^{0.215}_{3} = -0.88 \pm 0.08$; see Figure \ref{sed}) are inconsistent with this quasar being a blazar.
%}

\begin{table}
\centering
\caption{Flux densities and derived properties of P352--15 \label{photometry}}
\hspace*{-35pt}\begin{tabular}{lcc}
\hline
\hline

\hspace{0.4cm}{Telescope/Band} & Central $\lambda / \nu$ & Flux Density\\
\hline
Gemini--N/GMOS \textit{r}& 630 nm & $0.17 \pm 30$ $\mu$Jy\\
Pan--STARRS1 \textit{i}$_{P1}$ & 752 nm & $1.8 \pm 0.5$ $\mu$Jy\\
Pan--STARRS1 \textit{z}$_{P1}$ & 866 nm & $11.8 \pm 0.7$ $\mu$Jy\\
Pan--STARRS1 \textit{y}$_{P1}$ & 962 nm & $13.6 \pm 1.5$ $\mu$Jy\\
Magellan/FourStar \textit{J} &  1242 nm & $14.6 \pm 0.4$ $\mu$Jy\\
ALMA Band--7 & 290 GHz &  $0.34 \pm 0.04$ mJy\\
NOEMA Band--1 & 100 GHz & $0.10 \pm 0.01$ mJy\\
VLA--S & 3 GHz &   $8.20 \pm 0.25$ mJy\\
VLA--L & 1.4 GHz & $14.9 \pm 0.7$ mJy\\
GMRT Band--235\,MHz & 215 MHz &   $88 \pm 7$ mJy\\

\hline
\hline
\multicolumn{3}{c}{Rest Frame Luminosity \textit{L} ($L_\odot$)} \\
\hline
$L_{\mathrm{5\, GHz}}$ & &  $10^{10.28 \pm 0.01}$ \\
$L_{\mathrm{4400\,\angstrom}}$ & & $10^{12.37 \pm 0.11}$ \\
$L_{\mathrm{2500\,\angstrom}}$ & &$10^{12.49 \pm 0.03}$ \\

\hline
\hline
\multicolumn{3}{c}{Radio Loudness} \\
\hline
$R_{4400}$ & & $1100 \pm 280$\\
$R_{2500}$ & &  $1470^{+110}_{-100}$\\
\hline
\end{tabular}
\end{table}

\subsection{Modeling the mm Emission} \label{mm-text}
In this work we present two mm-continuum measurements of P352--15 from ALMA 290~GHz and NOEMA 100 GHz observations with S/N$>8$ (see Figure~\ref{cont}), corresponding to the quasar's rest-frame frequencies of 1981~GHz and 683~GHz, respectively. 
The flux densities at these frequencies are consistent with being at the Rayleigh–Jeans tail (optical depth $\tau_{Dust} << 1$) of a modified blackbody function (MBB), thought to be a good representation of cold dust from the quasar host galaxy \citep[e.g.,][]{priddey_far-infrared-submillimetre_2001,beelen_350_2006,da_cunha_effect_2013, leipski_spectral_2014,venemans_bright_2016,venemans_dust_2018}. The Planck function is:

\begin{equation}
B_{\nu_{rest}} (T_{Dust,z}) = \frac{2 h \nu_{rest}^3 / c^2}{e^{(h\,  \nu_{rest} / k\, T_{Dust,z})} - 1}
\end{equation}
where $T_{Dust,z}$ is the dust temperature at a given redshift. Typical dust parameters of the MBB found for host galaxies of radio-quiet quasars are $T_{Dust} = 47$ K and dust emissivity spectral index $\beta = 1.6$ \citep[e.g.,][]{beelen_350_2006,venemans_bright_2016}. In the following analysis we will assume these best-fit parameters and check whether they represent the properties of this quasar known to have strong radio emission. Later on, we will also explore whether different values of $T_{Dust}$ and $\beta$ can provide a better representation of the existing data of P352--15. 

\begin{figure*}[ht!]
\centering
\includegraphics[width=\textwidth]{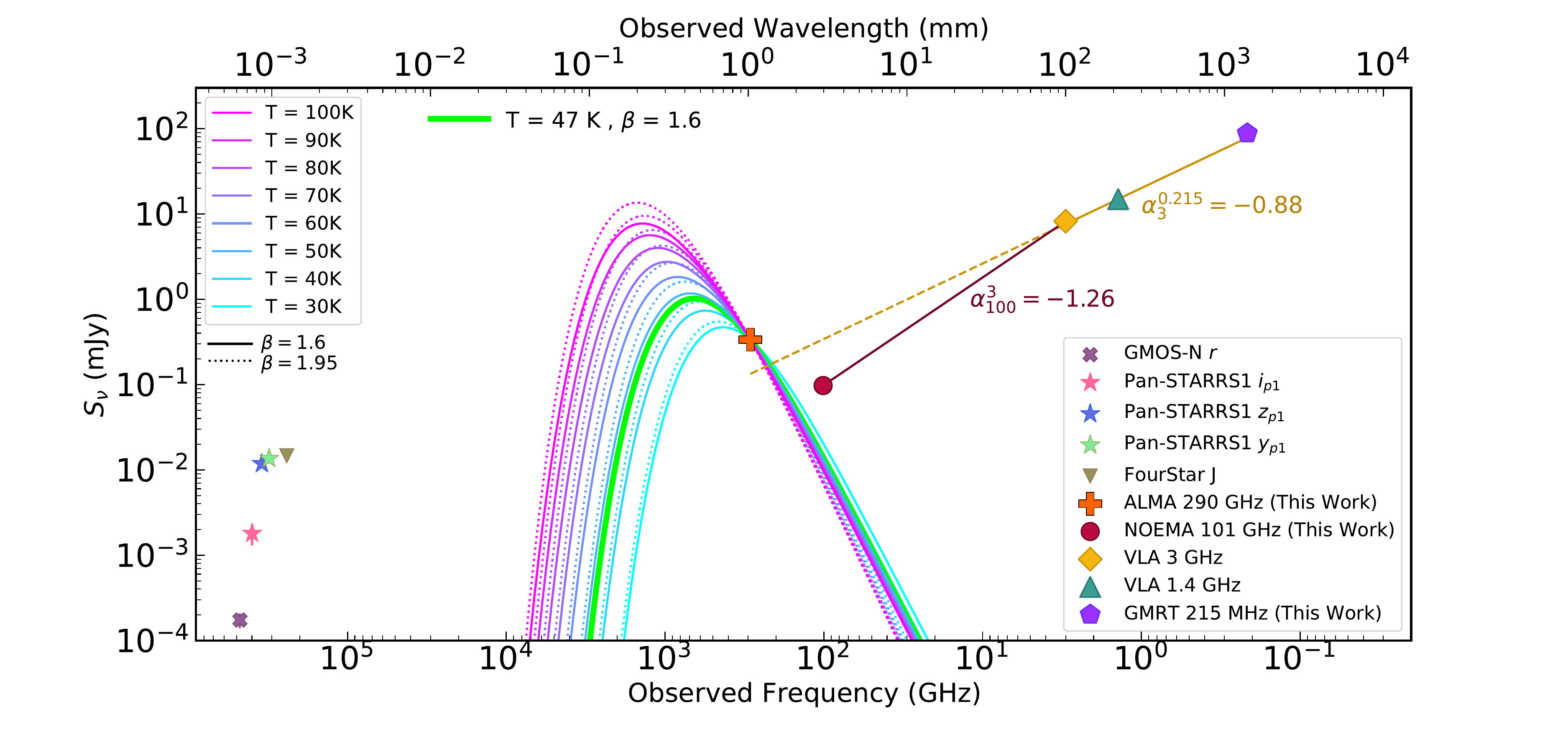}
\caption{Optical, millimeter and radio spectral energy distribution (SED) of P352--15 at $z\!=\!5.832$. The millimeter ALMA and NOEMA measurements are inconsistent with cold dust modeled as a modified black body at different dust temperatures and dust emissivity spectral indexes. The radio data at $215$\,MHz$-$3\,GHz are well described by synchrotron emission with a power law slope $\alpha \!= \!-0.88$. However, extrapolating that power law (dashed line) would be inconsistent with the millimeter data. To explain this SED, the synchrotron emission must steepen or break at high frequencies. We note that the error bars from the measurements are small compared to the scale of the figure and thus are not shown in the image.
  }
 \label{sed}
\end{figure*}

The predicted observed flux density from dust heating can be calculated with the MBB following \citet{novak_alma_2019}:

\begin{multline}
    S_{obs} = f_{\rm CMB} \ \left[1 + z\right] \ D_L^{-2} \  \kappa_{\nu_{rest}}(\beta) \\
     \times M_{Dust} \ B_{\nu_{rest}}(T_{Dust,z})
     \label{mod-bb}
\end{multline}

Here, $f_{\rm CMB}$ is a correction against the Cosmic Microwave Background (CMB) contrast, D$_L$ is the luminosity distance, $\kappa_{\nu,\,{rest}}(\beta)$ is the dust mass opacity, which depends on the dust emissivity spectral index $\beta$, and \Mdust is the dust mass. All values are given in SI units. 

The $f_{\rm CMB}$ term corrects for CMB contrast and heating effects, which can be significant at these high redshifts as we are seeing fluxes at frequencies close to the peak of the CMB \citep[see][]{da_cunha_effect_2013}. However, we do not include a CMB heating correction in Equation \ref{mod-bb} as this effect is still not significant for $T_{Dust,0}=30-100$\,K at $z=5.832$ (including CMB heating changes the temperature in that range by $<0.3$\,K). Thus, $f_{\rm CMB}$ in Equation \ref{mod-bb} is only the correcting factor for CMB contrast where the blackbody calculated with the temperature of the CMB at the redshift of the source is divided by the blackbody at the dust temperature: 
 
\begin{equation}
    f_{\rm CMB} = 1 - \frac{B_{\nu_{rest}} \ (T_{CMB,z})}{B_{\nu_{rest}} \ (T_{Dust,z})}
    \label{fcmb}
\end{equation}
\vspace{-5pt}

For $z=5.832$ and $T_{Dust}=47\,K$, $f_{\rm CMB}$ corresponds to 0.96 and 0.79 for the ALMA 290 GHz and NOEMA 100 GHz continuum points, respectively. 

We assume for $\kappa_{\nu,\,{rest}}(\beta)$ the relation derived in the submillimeter at 850\,$\mu$m in \citet{dunne_scuba_2000} and \citet{james_scuba_2002}: 
\begin{equation}
    \kappa_{\nu_{rest}}(\beta) =  0.077 \ (\nu\,_{rest} \ / \ 352 \ \mathrm{GHz})^\beta \ \mathrm{m^2~kg^{-1}}
    \label{kappa}
\end{equation}

We scale the MBB (Equation \ref{mod-bb})  to match the observed ALMA 290 GHz continuum and obtain a dust mass value of $(0.36\pm 0.04)\times 10^8$ $M_\odot$. We note that if instead of using Equation \ref{kappa} we assume $\kappa_{\nu,\,_{rest}} = 2.64~ \mathrm{m^2~kg^{-1}}$ at $\nu_{rest} = c/(125~ \mu$m) as presented in \citet{dunne_census_2003}, the value of \Mdust would decrease by a factor of 1.6. 

Using the calculated dust mass value and Equation \ref{mod-bb}, we can model the MBB at different frequencies. This model predicts a flux density of 0.0139 mJy at 100\,GHz, while our NOEMA measurement is 0.10 mJy. Therefore, these values are inconsistent by an order of magnitude and are 7.6$\sigma$ off from the observed NOEMA flux density. Since neither value set is able to fit the MBB function to both ALMA and NOEMA continuum measurements, we explore the different MBB functions that can be obtained when varying $T_{Dust}$ = 30 K -- 100 K, and $\beta = 1.6 $  and $\beta = 1.95$  as shown in Figure \ref{sed}. We also try to find a $\beta$ value that could fit both of our mm-continuum points. 
This results in $\beta = -0.7$, however, such a negative dust emissivity index has never been reported for dust in galaxies \citep[see e.g.,][]{dunne_scuba_2000,beelen_350_2006}. 

It is evident that none of these models can reproduce our data, implying that in this system the mm-emission is not only due to cold dust but it must be affected by an additional source, the main suspect being the strong synchrotron emission of the quasar, as discussed below. \\

%====================================================
%============ Discussion =======================
%====================================================
\section{Discussion} \label{discussion}
In this section we discuss the implications of our results for the intriguing SED of the radio-loud quasar P352--15. We first describe the possible influence of synchrotron emission up to higher frequencies, which are usually dominated by dust continuum. Next, we report the derived far infrared (FIR) properties for the quasar based on the mm continuum measured with ALMA at 290 GHz, assuming that this measurement is not contaminated by synchrotron radiation.

\subsection{Effects of the synchrotron emission on the mm continuum measurements.} \label{effects}
From the observations described in this work, we see that the radio spectrum is clearly defined by the power-law slope of $\alpha^{0.215}_{3} = -0.88 \pm 0.08$. However, the continuum emission at millimeter wavelengths is not well-matched to a blackbody fit (Figure \ref{sed}). Given the strong radio emission in this quasar, a possible explanation would be that the millimeter emission is highly affected by synchrotron radiation instead of being only due to cold dust \citep[e.g.,][]{weis_laboca_2008}

In order to assess this possibility, we use the synchrotron power-law fit derived in Section \ref{model-radio} and extrapolate it to the mm data points. We would expect the extrapolated flux at 100\,GHz to be above the measured NOEMA continuum, but the flux density at 290\,GHz would be below the measured ALMA continuum (see dashed line in Figure~\ref{sed}). When evaluating the slope between the VLA 3\,GHz and NOEMA 100\,GHz flux density values, the observed  spectral index is much steeper ($\alpha_{100}^{3} = - 1.26 \pm 0.03$) than $\alpha^{0.215}_{3}$. 
This means that extrapolating the synchrotron power-law we would have expected a NOEMA detection about three times brighter than what we measure. Therefore, it is very likely that the 100\,GHz emission in this quasar is dominated by synchrotron emission. Further measurements between 3--100\,GHz are required to pinpoint the location of an expected spectral break. 

Extrapolating the synchrotron power-law ($\alpha^{0.215}_{3} = -0.88$) all the way to 290\,GHz results in 0.13\,mJy. This implies that the synchrotron contribution in the ALMA measurement could be up to 40\% of the measured value. However, this is a strict upper limit given that we know there is a break in the power-law spectrum between 3 and 100\,GHz (Figure \ref{sed}). If we extrapolate to 290\,GHz using the $\alpha_{100}^{3} = - 1.26$ power-law, we expect a synchrotron contribution of only 8\% in the ALMA continuum measurement.  

%\edit1{
Due to the significant synchrotron contribution for the NOEMA 100\,GHz observations, and a smaller influence on the ALMA 290\,GHz, we explore an overall fit of the synchrotron emission and the cold dust, to model both components of the SED simultaneously from the radio to mm regime. For this purpose, we tried to fit the modified black body function (MBB) and a simple power-law function, but this model cannot reproduce our data. We then performed a joint fit using an MBB and a broken power-law and fix the break frequency at 3 GHz, that is the limit of our observations in radio. The resulting model is over-fitted given the few data points and the fitted parameters do not yield trusting values. Since more observations at complementary frequencies are needed to determine the break frequency, we continue to constrain the two components in mm and radio separately.
%}

\subsection{FIR properties of P352--15}
We use the ALMA continuum measurement to estimate the dust properties of the host galaxy, assuming that the synchrotron emission at this frequency is minimal. However, we caution the reader that these measurements should be considered as upper limits given that a non-negligible contribution from the synchrotron emission is possible. Additional measurements covering the 3~GHz to 100~GHz gap are required to better disentangle the synchrotron emission at this frequency. 

From the \Mdust derived in Section \ref{mm-text} along with the dust parameters T = 47 K and $\beta = 1.6$, we integrate over the SED and obtain infrared luminosity constraints. We calculate the FIR luminosity by integrating over rest-frame 42.5--122.5$\,\mu$m and get \Lfir $ = (0.89 \pm 0.10) \times 10^{12}$ $L_\odot$. Integrating over the rest-frame wavelengths 8--1000$\,\mu$m, we obtain the Total Infrared (TIR) luminosity as \Ltir $ = (1.26 \pm 0.15) \times 10^{12}$~$L_\odot$.  
We derive a SFR from the \Ltir following the relation in \citet{kennicutt_star_2012}  scaled to the \citet{chabrier_galactic_2003} initial mass function (IMF; this results in a SFR 1.7 times smaller than assuming the \citet{salpeter_luminosity_1955} IMF), obtaining SFR$_{\mathrm{TIR}} =  (110.0 \pm 13.0)$ $M_\odot$~yr$^{-1}$. This value is $\sim 15\%$ smaller than the resulting from calibrations in \citet{kennicutt_star_1998}. The uncertainty in SFR$_{\mathrm{TIR}}$ can be larger because we cannot constrain the dust parameters for temperature T, and emissivity index $\beta$ (see Figure~\ref{sed}).

Using the equations for calculating line luminosities from \cii\ measurements presented by \citet{carilli_cool_2013}, we find the areal integrated source brightness temperature $L\mathrm{'}_{\mathrm{[C\, \sc{II}]}} = (5.6\pm 0.9) \times 10^9$ K km~s$^{-1}$ pc$^2.$ The \cii\ line luminosity, which is commonly used to compare luminosities to the underlying continuum, is $L_{\mathrm{[C\, \sc{II}]}} = (1.23\pm 0.20) \times 10^9$~$L_\odot$. We use the relation in \citet{de_looze_applicability_2014} to calculate the \cii-derived SFR:

\begin{equation}
    \mathrm{log} \, \mathrm{SFR}_{\mathrm{[C\, \sc{II}]}} = -6.09 + 0.90 \times  \mathrm{log} \ \textit{L}_{\mathrm{[C\, \sc{II}]}}
\end{equation}
Taking into account the systematic uncertainty in this relation of a factor $\sim$2.3, we find SFR$_{\mathrm{[C\, \sc{II}]}} = (54-285)$~$M_\odot$~yr$^{-1}$, which is consistent with the SFR inferred from the TIR luminosity, as indicated in Table~\ref{lums}. 

\begin{table}
\begin{center}
\caption{Derived FIR properties for P352--15 \label{lums}}
\begin{tabular}{lccccc}
\hline
\hline
\Mdust\tablenotemark{$\ast$} & & & & &  $(0.36\pm 0.04)\times 10^8$ $M_\odot$ \\
\Lfir\tablenotemark{$\ast$} & & & & &  $(0.89 \pm 0.10) \times 10^{12}$ $L_\odot$  \\
\Ltir\tablenotemark{$\ast$}  & & & & &  $(1.26 \pm 0.15) \times 10^{12}$ $L_\odot$ \\
SFR$_{\mathrm{TIR}}$\tablenotemark{$\ast$} & & & & & $(110.0 \pm 13.0)$ $M_\odot$~yr$^{-1}$ \\
$L\mathrm{'}_{\mathrm{[C\, \sc{II}]}}$ & & & & & ($5.6\pm 0.9) \times 10^9$ K km~s$^{-1}$ pc$^2$ \\
$L_{\mathrm{[C\, \sc{II}]}}$  & & & & & ($1.23\pm 0.20) \times 10^9$ $L_\odot$ \\
SFR$_{\mathrm{[C\, \sc{II}]}}$\tablenotemark{\dag} & & & & & ($54-285$) $M_\odot$~yr$^{-1}$ \\
$L_{\mathrm{[C\, \sc{II}]}}$/\Lfir  & & & & & $(1.38\pm 0.23)\times 10^{-3}$ \\
$L_{\mathrm{CO(6-5)}}$ & & & & & $ < 0.5 \times 10^9$ $L_\odot$\\
\hline
\end{tabular}
\end{center}
\tablenotetext{$$\ast$$}{These values are calculated assuming a MBB matched to the ALMA continuum with $T_{Dust}=47$ K and $\beta = 1.6$. We caution that these measurements might be affected by synchrotron emission.}
\tablenotetext{\dag}{Calculation using the SFR to luminosity relation derived in \citet{de_looze_applicability_2014}.}
\end{table}

%====================================================
%============ Comparison =======================
%====================================================
\begin{figure*}[ht!]
\centering
\includegraphics[width=0.55\linewidth]{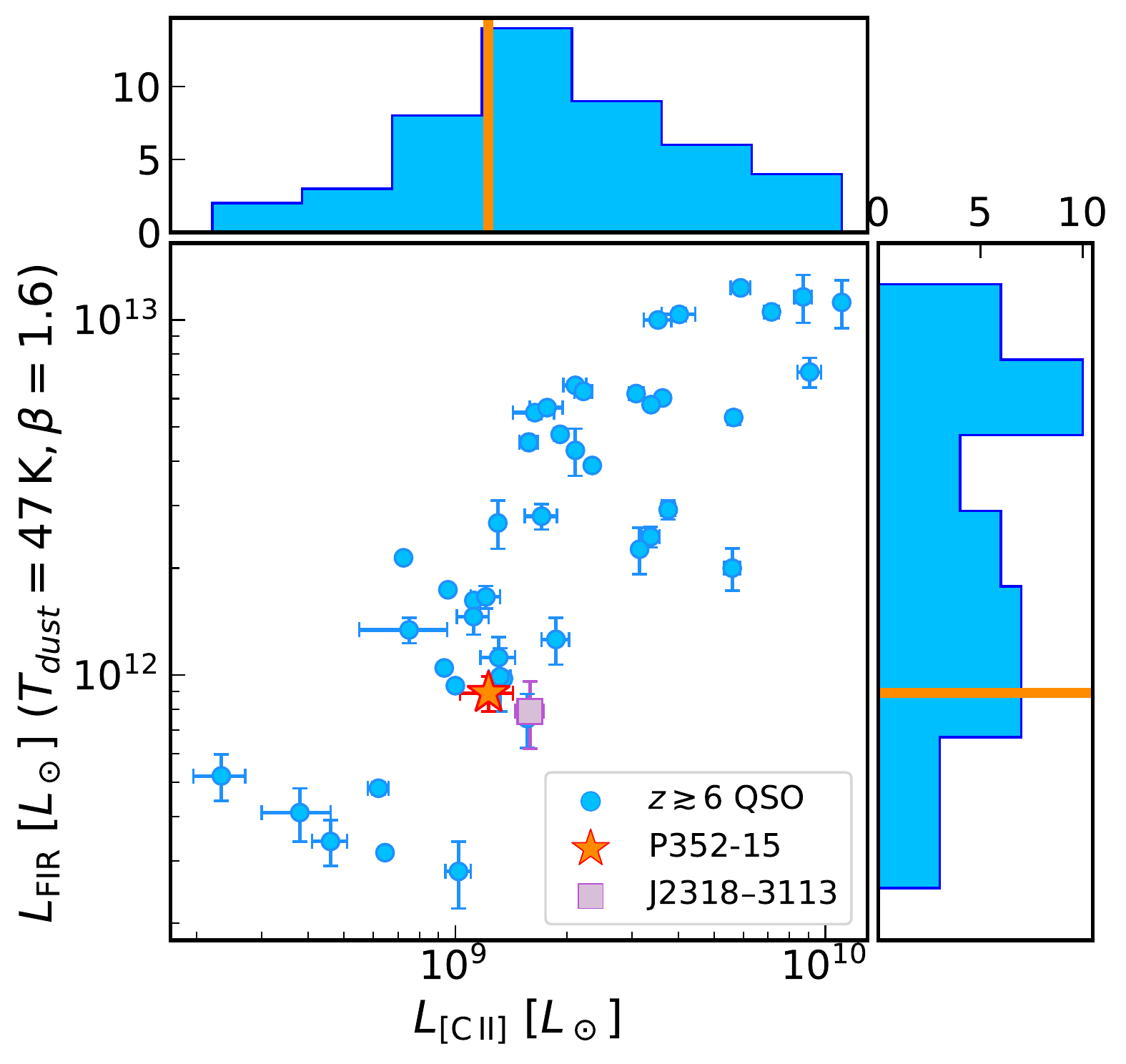}
\hfill
\includegraphics[width=0.44\linewidth]{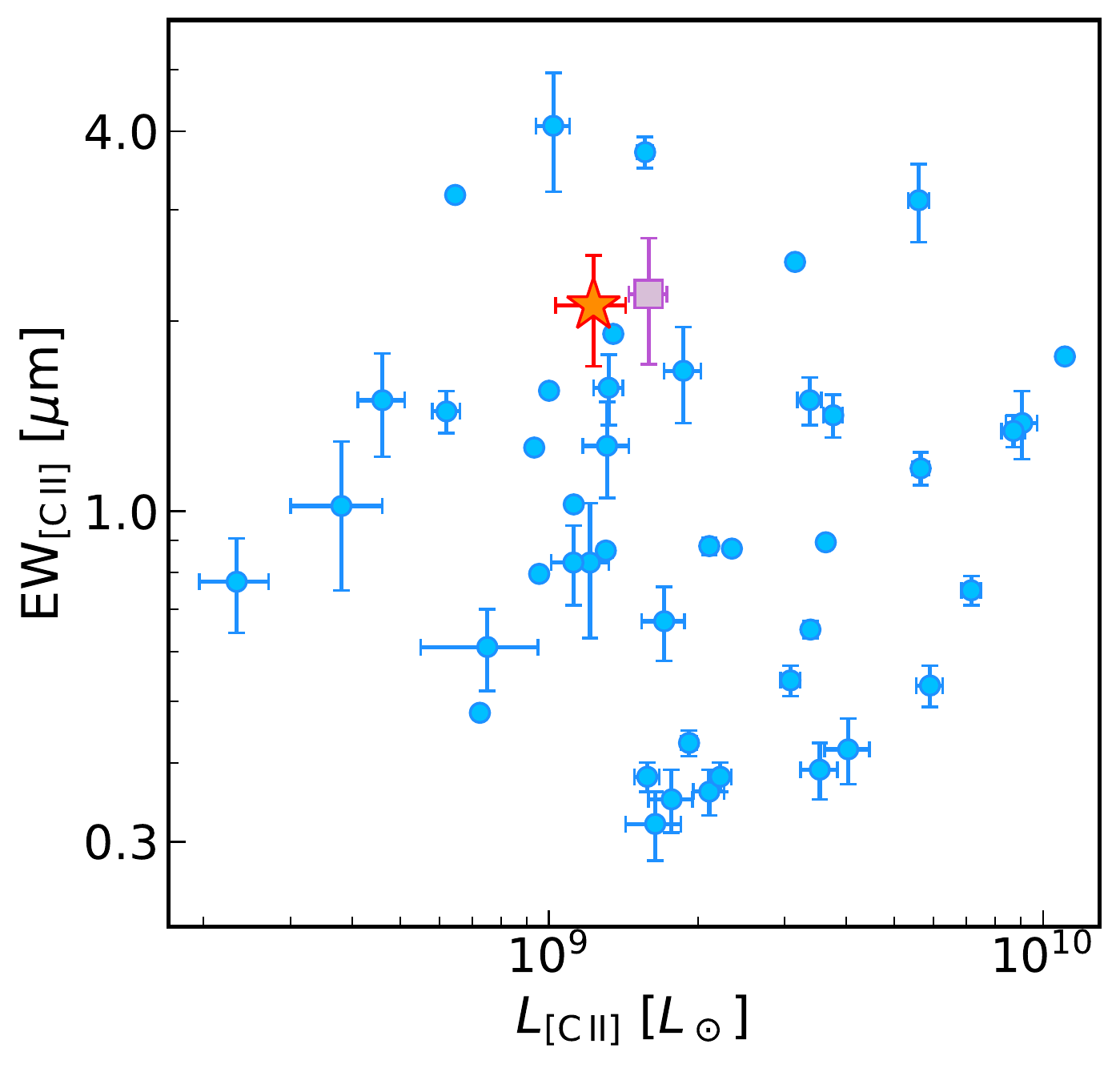}
\caption{\textit{Left:} Relationship between the FIR luminosity (assuming is dominated by cold dust with $T_{Dust}=47$ K and $\beta = 1.6$), and the \cii\ luminosity reported in 45 $z \gtrsim 6$ quasars from \citet{decarli_alma_2018,izumi_subaru_2018,eilers_detecting_2020,eilers_erratum_2021,venemans_kiloparsec-scale_2020}. The derived properties of P352--15 are within typical values for \Lfir\ and $L_{\mathrm{[C\, \sc{II}]}}$ in context with the literature. \textit{Right:} Comparison between the \cii\ equivalent width and the \cii\ luminosity using the same sample of $z \gtrsim 6$ quasars. P352--15's EW$_\mathrm{[C\, \sc{II}]}$ value is at the high end of the distribution found in the literature. The two panels also highlight J2318--3113, the only other radio-loud quasar in the $z \gtrsim$ 6 sample with available information for comparison. Both radio-loud quasars have very similar \cii\ and FIR properties.}
\label{comp_fig}
\end{figure*}

\subsection{Comparison with the Literature}\label{compare}

Here we place the \cii\ and dust emission properties for P352--15 in the context of other quasars at  $z \gtrsim 6$ in the literature. In Figure \ref{comp_fig}, we compile the properties of $z\gtrsim 6$ quasars with continuum measurements of S/N$>3$ as reported by \citealt{decarli_alma_2018, izumi_subaru_2018, eilers_detecting_2020,eilers_erratum_2021, venemans_kiloparsec-scale_2020}. We note that all the literature quasars are radio-quiet with the only exception of J2318--3113, whose \cii\ properties have been studied by \cite{decarli_alma_2018} and  \cite{venemans_kiloparsec-scale_2020}, but it was only recently identified as radio-loud (with $R_{4400}\sim 70$) by \cite{ighina_radio_2021}. Both radio-loud quasars with \cii\ information, P352--15 and J2318--3113, are highlighted in Figure~\ref{comp_fig}.

P352--15 has a \cii\ line width of $440\pm 80$ km~s$^{-1}$, which is broader but within one standard deviation than the mean of the distribution from $z \gtrsim 6$ quasars in the literature \citep[350 $\pm$ 125 km/s;][]{decarli_alma_2018, venemans_kiloparsec-scale_2020}. The \cii\ luminosity we calculate for P352--15 is $(1.23\pm 0.20) \times 10^9$ $L_\odot$, which is smaller than the median of the distribution but it is still within the typical values of $(1-5)\times 10^9$\,$L_\odot$ for quasars at $ z \gtrsim 6$ (see Figure \ref{comp_fig} and e.g., \citealt{mazzucchelli_physical_2017,venemans_copious_2017,novak_alma_2019}).

As mentioned in Section \ref{sed}, the dust properties for quasar P352--15 cannot be fully constrained given the strong synchrotron contamination at millimeter wavelengths. However, we see that the upper limit estimate for \Lfir\ $\sim10^{12}$ $L_{\odot}$ overlaps with those of quasar host galaxies at $z \gtrsim 6$ in the literature, although it is smaller than the median value of the population (Figure~\ref{comp_fig}). Similarly, the SFR implied by the total infrared emission is consistent with the one we derive from the \cii\ line, and with previous studies of quasars at these redshifts \citep[e.g.,][]{mazzucchelli_physical_2017,eilers_detecting_2020,eilers_erratum_2021}.

A widely used diagnostic to study properties of the interstellar medium (ISM) is the \cii-to-FIR luminosity ratio. Local galaxies ($z <$ 1) show a decrease in this ratio towards higher FIR luminosities (referred to as the `\cii\ deficit'), driven by the presence of mechanisms provoking dust heating in the ISM \citep[e.g.,][]{malhotra_far-infrared_2001,farrah_far-infrared_2013,sargsyan_star_2014,diaz-santos_aherschelpacs_2017}. The high-redshift ($z \gtrsim 6$) sample has a large scatter for this ratio but our calculated log(L$_{\mathrm{[C\, \sc{II}]}}$ / \Lfir) =$-2.86$ is comparable to local ultra-luminous infrared galaxies (ULIRGs), and is consistent with previously found measurements of  $z >$ 6 quasars (see trend in Figure \ref{comp_fig}, left) \citep[e.g.,][]{venemans_detection_2012,venemans_copious_2017,banados_bright_2015,willott_star_2015,mazzucchelli_physical_2017,decarli_alma_2018}. 

A useful quantity to investigate is the \cii\ equivalent width (EW), which has the advantage that it does not depend on the shape of the dust SED: 
\begin{equation}
    \frac{\mathrm{EW}}{[\mu\mathrm{m}]} = 1000\ \frac{S \Delta v \mathrm{[Jy\ km~s^{-1}]}}{S_{\nu,0} (\mathrm{cont}) \mathrm{[mJy]}}\ \frac{\lambda_0 [\mu\mathrm{m}]}{c \mathrm{[km~s^{-1}]}}\\
\end{equation}
Here, $S \Delta v$ is the velocity-integrated line, $S_{\nu,0} (\mathrm{cont})$ is the observed continuum flux density at the rest-frame frequency of the line (\cii\ in this case), $\lambda_0$ is the rest-wavelength which for \cii\ is 157.74 $\mu$m, and $c$ is the speed of light. Our computed value for P352--15 is EW$_\mathrm{[C\, \sc{II}]} = 2.12 \pm 0.42\  \mu$m, placing P352--15's EW$_\mathrm{[C\, \sc{II}]}$ at the high end of the distribution  but with several other hosts of radio-quiet quasars having comparable or larger EW$_\mathrm{[C\, \sc{II}]}$ (see right panel in Figure \ref{comp_fig}). We conclude that the  EW$_\mathrm{[C\, \sc{II}]}$ of P352--15 is consistent with what is expected from the properties of radio-quiet quasars, and it is therefore likely that the continuum underlying the \cii\ emission is indeed not strongly influenced by the radio emission (see \S \ref{effects}).

It is worth noting that the only two $z\gtrsim 6$ radio-loud quasars studied in their rest-frame FIR and \cii\ properties lie in a very similar position in the parameter space of Figure \ref{comp_fig}. Nevertheless, a systematic mm study of a larger sample of radio quasars at high redshifts is necessary to be able to make statistical comparisons between the black hole/host galaxy properties of radio-loud and radio-quiet sources.

%====================================================
%============ Conclusion =======================
%====================================================
\section{Summary and Conclusions}\label{conclusion}
We present and analyze millimeter and radio observations of the powerful radio bright quasar P352--15. Our major findings are summarized as follows.
\begin{itemize}

\item The ALMA observations of the \cii\ line emission from this quasar resulted in a systemic redshift calculation of $z\!=\!5.832 \pm 0.001$. We find a \cii\ line width of $440\pm 80$ km~s$^{-1}$. The \cii\ luminosity is $1.23\times 10^{9}$ $L_{\odot}$, comparable to values reported in other high-redshift quasars in the literature. The derived star formation rate from the \cii\ emission is SFR$_{\mathrm{[C\, \sc{II}]}}$ = ($54-285$) $M_\odot$~yr$^{-1}$. Finally, the ALMA observations revealed an underlying unresolved continuum source at 290~GHz with a flux density of $0.34 \pm 0.04$\,mJy (see Figures \ref{cii}--\ref{cont}).

\item We do not detect the CO\,(6--5) emission line in our NOEMA 100 GHz observation. The derived $3\sigma$ upper limit is $<$0.35 mJy corresponding to a luminosity $L_{\mathrm{CO(6-5)}} < 0.5 \times 10^9$ $L_\odot$ (assuming the same FWHM as measured for the \cii\ line). Continuum emission is detected with NOEMA at 100~GHz with a flux density of $0.10 \pm 0.01$ mJy (see Figure~\ref{cont}). 

\item The high S/N GMRT 215 MHz observations of the quasar allow us to improve the constraints on the radio emission at low frequencies (c.f., \citealt{banados_powerful_2018}). The synchrotron radiation can be well modeled by a simple power-law with spectral index $\alpha^{0.215}_{3} = -0.88 \pm 0.08$. This resulted in updating the calculation of the quasar's radio-loudness parameter $R_{4400} = 1100 \pm 280$ and $R_{2500} = 1470^{+110}_{-100}$, confirming its prominent radio power (see Figure~\ref{sed}). 

\item We study the dust emission of the quasar host galaxy by using the typical approach of modeling a modified black body function (MBB) to our two mm-continuum measurements from ALMA 290 GHz and NOEMA 100 GHz. However, the NOEMA measurement is an order of magnitude brighter than expected from the MBB scaled to the ALMA emission using the typical $T_{dust}=47\,K$ and $\beta=1.6$, and cannot be modeled by any other dust parameters commonly used in the literature. Therefore we hypothesize that the 100~GHz continuum emission is dominated by synchrotron radiation. However, the emission is three times dimmer than the expected emission by extrapolating the trend at lower frequencies. This implies a break in the synchrotron spectrum (see~Figure~\ref{sed}).

\item From the ALMA 290 GHz continuum observations we derive the FIR properties of the quasar, assuming this data point is not significantly contaminated by synchrotron emission (see discussion in Section~\ref{effects}). Assuming a MBB with typical parameters ($T_{dust}=47\,K$ and $\beta=1.6$), we derive a dust mass and FIR and TIR luminosities. These quantities are reported in Table~\ref{lums}. The star formation rate derived from the \Ltir is SFR$_{\mathrm{TIR}} = (110.0 \pm 13.0)$ $M_\odot$~yr$^{-1}$, which is consistent with the calculated SFR from the \cii\ emission.

\item This quasar lies within typical estimates of \cii\ FWHM, EW$_{\mathrm{[C\, \sc{II}]}}$, $L_{\mathrm{[C\, \sc{II}]}}$, and \Lfir\ compared to quasars at $z \gtrsim 6$ in the literature. We highlight that the derived properties of P352--15 are very similar to quasar J2318--3113, which is the only other radio-loud quasar with \cii\ and rest-frame FIR studies (see~Figure~\ref{comp_fig}). A more comprehensive study of high-redshift radio quasars in the millimeter is required for statistical comparisons.

\end{itemize}

The extreme radio brightness of the quasar P352--15 makes it an ideal laboratory to study the effects of powerful radio jets on the formation of supermassive black holes and galaxies during the first Gyr of cosmic time. This quasar presents an intriguing SED (Figure \ref{sed}) and leaves more open questions that make it an excellent target for follow-up observations. For example, to evaluate the turnover frequency of the synchrotron spectrum, observations filling the gap between the existing 3 and 100\,GHz measurements are required (as shown in Figure \ref{sed}). If the slope steepens at observed frequencies higher than 3 GHz, we can assume the ALMA observations sample the dust emission while the NOEMA emission is dominated by synchrotron. This way it would be possible to disentangle the synchrotron from dust emission. Furthermore, at these high frequencies the relativistic electrons dissipate much faster assuming that synchrotron radiation is the main dissipation mechanism, causing a steepening of the synchrotron spectrum \citep{scheuer_radio_1968,pacholczyk_radio_1970,condon_essential_2016}. Therefore, by identifying the turnover frequency we would be able to gauge the age of the radio jet  \citep[e.g.,][]{myers_synchrotron_1985,carilli_multifrequency_1991}. Finally, our current data marginally resolves the \cii\ emission (Figure \ref{cii}); higher resolution observations could reveal potential signatures of on-going  merger or outflow in this quasar. All of these open questions can be addressed with future observations from existing facilities such as ALMA, NOEMA, and the VLA. \\

%=====================================================
%============ ACKNOWLEDGEMENTS =====================
%=====================================================
\acknowledgements
We thank Charl\`ene Lef\`evre at IRAM for the great help and guidance reducing the NOEMA data used in this work.
S.R.R. acknowledges financial support from the International Max Planck Research School for Astronomy and Cosmic Physics at the University of Heidelberg (IMPRS--HD). The work of T.C. was carried out at the Jet Propulsion Laboratory, California Institute of Technology, under a contract with NASA. A.C.E. acknowledges support by NASA through the NASA Hubble Fellowship grant $\#$HF2-51434 awarded by the Space Telescope Science Institute, which is operated by the Association of Universities for Research in Astronomy, Inc., for NASA, under contract NAS5-26555. The National Radio Astronomy Observatory is a facility of the National Science Foundation operated under cooperative agreement by Associated Universities, Inc.

This paper makes use of ALMA data in program ADS/JAO.ALMA$\#$ 2019.1.00840.S. The Joint ALMA Observatory is operated by ESO, AUI/NRAO and NAOJ. This work is based on observations carried out under project number W18EG with the IRAM NOEMA Interferometer. IRAM is supported by INSU/CNRS (France), MPG (Germany), and IGN (Spain). This paper uses GMRT observations from program ddtC007. We thank the staff of the GMRT that made these observations possible. GMRT is run by the National Centre for Radio Astrophysics of the Tata Institute of Fundamental Research.

\facility{ALMA,
        IRAM:NOEMA,
        GMRT}
\software{Astropy \citep{the_astropy_collaboration_astropy_2013},
          CASA \citep{mcmullin_casa_2007},
          Interferopy (\url{https://github.com/mladenovak/interferopy}),
          Matplotlib \citep{hunter_matplotlib_2007},
          Numpy \citep{harris_array_2020}}

%=====================================================
%============ References ===========================
%=====================================================
\bibliographystyle{yahapj}
\bibliography{references.bib}
\end{document}